\documentclass{article}
\usepackage{spconf,amsmath,graphicx,hyperref}

\usepackage{algorithm}
\usepackage{algorithmicx}
\usepackage{algpseudocode}
\usepackage{amssymb}
\usepackage{amsfonts}
\usepackage{arydshln}
\usepackage{bm}
\usepackage{color}
\usepackage{multirow}
\usepackage{soul}
\usepackage{float}
\usepackage{mathtools}
\usepackage[utf8]{inputenc}
\usepackage[english]{babel}
\usepackage{amsthm}
\usepackage[nopostdot,nogroupskip,nonumberlist]{glossaries}
\usepackage{cite}
\usepackage{booktabs}
\usepackage{subfig}
\usepackage[T1]{fontenc}
\usepackage{url}
\usepackage{ifthen}

\usepackage[font=small,skip=6pt]{caption} 

\usepackage{etoolbox}
\AtBeginEnvironment{equation}{\useshortskip}
\AtEndEnvironment{equation}{\useshortskip}
\AtBeginEnvironment{align}{\useshortskip}
\AtEndEnvironment{align}{\useshortskip}
\AtBeginEnvironment{subequations}{\useshortskip}
\AtEndEnvironment{subequations}{\useshortskip}
\newcommand{\useshortskip}{%
	\setlength{\abovedisplayskip}{3pt plus 2pt minus 2pt}
	\setlength{\belowdisplayskip}{3pt plus 2pt minus 2pt}}

\hypersetup{colorlinks=true, allcolors=blue}
\glsdisablehyper 
\newacronym{ISAC}{ISAC}{integrated sensing and communication}
\newacronym{MIMO}{MIMO}{multiple-input-multiple-output}
\newacronym{BS}{BS}{base station}
\newacronym{iid}{i.i.d.}{independent and identically distributed}
\newacronym{RIS}{RIS}{reconfigurable intelligent surface}
\newacronym{UPA}{UPA}{uniform planar array}
\newacronym{LoS}{LoS}{line-of-sight}
\newacronym{NLoS}{NLoS}{non-line-of-sight}
\newacronym{RCS}{RCS}{radar cross-section}
\newacronym{SINR}{SINR}{signal-to-interference-plus-noise ratio}
\newacronym{SDR}{SDR}{semidefinite relaxation}
\newacronym{FFBF}{FFBF}{far-field beamforming}
\newacronym{NCCS}{NCCS}{no cross-correlation suppression}
\newacronym{AO}{AO}{alternating optimization}
\newacronym{MI}{MI}{mutual information}
\newacronym{SDP}{SDP}{semidefinite program}

\DeclarePairedDelimiter\abs{\lvert}{\rvert}%
\DeclarePairedDelimiter\norm{\lVert}{\rVert}%

\newcommand\myvec[1]{\bm{#1}}
\newcommand\mymat[1]{\mathbf{#1}}
\newcommand\mymatb[1]{\bar{\mathbf{#1}}}

\newcommand\mymath[1]{\mathbf{\hat{#1}}}

\newcommand\myvech[1]{\hat{\bm{#1}}}
\newcommand\myvecb[1]{\bar{\bm{#1}}}
\newcommand\myexpect[1]{\mathbb{E}\left\{#1\right\}}
\makeatletter
\let\oldabs\abs
\def\abs{\@ifstar{\oldabs}{\oldabs*}}
\let\oldnorm\norm
\def\norm{\@ifstar{\oldnorm}{\oldnorm*}}
\makeatother

\makeatletter
\renewcommand{\fnum@figure}{Fig. \thefigure}
\makeatother

\newcommand{\diag}{\mbox{diag}}

\newtheorem{theorem}{Theorem}

\newtheorem{remark}{\it{Remark}}

\graphicspath{{figures/}}

\title{RIS-ASSISTED NEAR-FIELD ISAC FOR MULTI-TARGET INDICATION IN NLOS SCENARIOS}

\name{Hang Ruan$^{1}$, Homa Nikbakht$^{1,2}$, Ruizhi Zhang$^{1,3}$, Honglei Chen$^{4}$, and Yonina C. Eldar$^{1}$
	\thanks{
		Emails: \{hang.ruan, yonina.eldar\}@weizmann.ac.il, homa@princeton.edu,
		zrz\_cdut@163.com, h.chen@mathworks.com}}

\address{
		$^{1}$Weizmann Institute of Science, $^{2}$Princeton University \\ 
		$^{3}$University of Electronic Science and Technology of China, $^{4}$MathWorks, Inc. 
}

\begin{document}
	\ninept
	\maketitle

	\begin{abstract}
		Enabling multi-target sensing in near-field integrated sensing and communication (ISAC) systems is a key challenge, particularly when line-of-sight paths are blocked. This paper proposes a beamforming framework that leverages a reconfigurable intelligent surface (RIS) to achieve multi-target indication. Our contribution is the extension of classic beampattern gain and inter-target cross-correlation metrics to the near-field, leveraging both angle and distance information to discriminate between multiple users and targets. We formulate a problem to maximize the worst-case sensing performance by jointly designing the beamforming at the base station  and the phase shifts at the RIS, while guaranteeing communication rates. The non-convex problem is solved via an efficient alternating optimization (AO) algorithm that utilizes semidefinite relaxation (SDR). Simulations demonstrate that our RIS-assisted framework enables high-resolution sensing of co-angle targets in blocked scenarios. 
	\end{abstract}
	\glsresetall
	
	\section{Introduction}
	\label{sec:intro}
	
	Future 6G networks require high-accuracy sensing alongside high-speed communication, for which \gls{ISAC} is a pivotal technology \cite{liu2022integrated}. By sharing spectrum and hardware, ISAC improves spectral efficiency, reduces costs, and creates synergy between the two functions \cite{liu2022integrated}.
	
	The trend towards large antenna arrays and high frequency pushes \gls{ISAC} into the near-field regime, where the spherical wavefront provides a distance dimension \cite{wang2023near,galappaththige2024near}. This critically enables the discrimination of co-angle targets, which is impossible in the far-field \cite{wang2023near,galappaththige2024near}.	
	Concurrently, \glspl{RIS} enhance \gls{ISAC} systems by creating virtual links to overcome signal blockages \cite{liu2023integrated}. While this has spurred research into \gls{RIS}-assisted near-field \gls{ISAC} \cite{dong2025semipassive, huang2024jointdesign, zhou2025nearfieldxl, xue2024nearfieldisac, zhu2024risassistednf}, these studies focus on single-target scenarios and cannot sense multiple targets simultaneously.

	Recognizing this limitation, a growing body of research has started to address the more challenging multi-user, multi-target \gls{ISAC} problem with \gls{RIS} assistance. These works have explored various performance metrics to manage the intricate trade-offs, such as the sensing \gls{SINR} \cite{zhang2024intelligent}, beam pattern gain \cite{zhang2025joint}, \gls{MI} \cite{yang2024risassisted}, and detection probability \cite{zhao2024dualfunctional, wang2024secure}.
	However, these works are fundamentally limited by their far-field assumptions, as their angle-only sensing metrics are insufficient for near-field scenarios where resolving targets in the same direction requires exploiting the distance dimension.
	
	To effectively harness near-field propagation for multi-target sensing, our prior work in \cite{ruan2025near} extended the robust metrics of transmit beampattern gain and inter-target cross-correlation to the near-field. This approach is offers two distinct advantages. First, unlike metrics such as sensing \gls{SINR}, \gls{MI}, or detection probability, our approach is agnostic to the targets' often unknown \glspl{RCS}. Second, the explicit suppression of inter-target cross-correlation is critical for accurate multi-target indication; without it, metrics based on \gls{SINR}, beampattern gain only or detection probability are insufficient for resolving multiple targets with classical \gls{MIMO} techniques \cite{stoica2007probing, meng2024multi}. 
	
	However, the framework in \cite{ruan2025near} was limited to \gls{LoS}-only scenarios. This paper bridges that critical gap by introducing a \gls{RIS} to overcome \gls{NLoS} blockages in a multi-user, multi-target near-field \gls{ISAC} system. Our proposed framework leverages the near-field's spherical wavefront to resolve targets in both angle and distance, even when obstructed.

	The main contributions of this work are as follows: 
	1) We formulate a joint transmission beamforming and \gls{RIS} response optimization problem for \gls{RIS}-assisted near-field \gls{ISAC} that accommodates multiple \gls{NLoS} users and targets.
	2) To enable sensing of multiple targets, we first extend classical far‐field beam‐pattern and cross‐correlation metrics to the near-field, 
	enabling the discrimination of objects at identical directions but different distances.
	3) We develop an efficient \gls{AO} algorithm to solve the complex non-convex problem. The sub-optimizations of both transmission beamforming and \gls{RIS} response are solved via the \gls{SDR} technique.
    4) We demonstrate through numerical simulations that our proposed RIS-assisted design overcomes blockages to achieve significant performance gains, enabling high-resolution sensing and communication in scenarios.
	
	
	\section{System and Signal Model}
	\label{sec:model}

	\subsection{System Model}
\begin{figure}[t!]
		\centering	
		\includegraphics[width=0.7\linewidth]{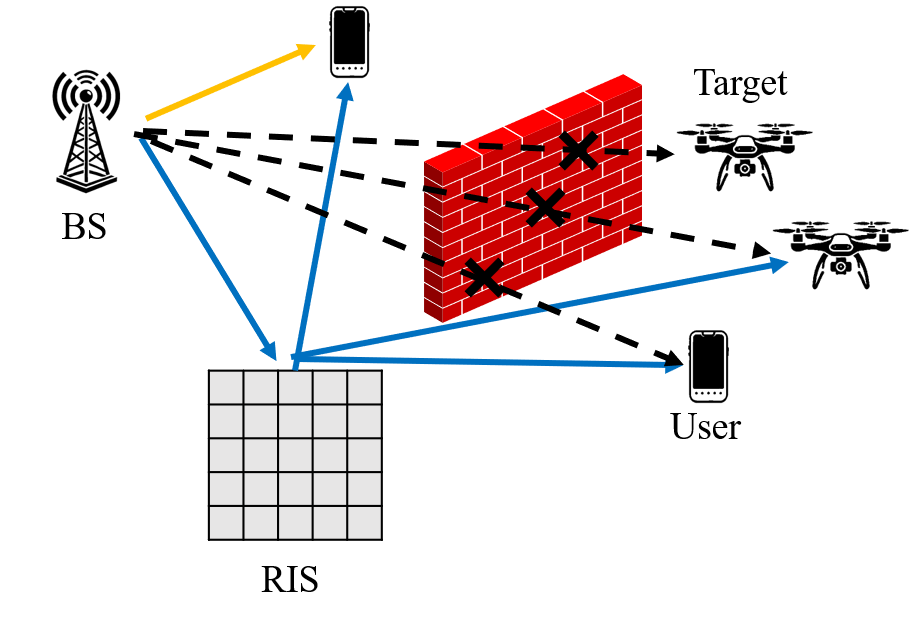} 
		\caption{Illustration of the RIS-assisted near-field ISAC.}
		\vspace{-0.5cm}
		\label{fig:scenario_ris}
	\end{figure}

	\vspace{-2ex}
	We consider a near-field mmWave \gls{ISAC} system (Fig.~\ref{fig:scenario_ris}) comprising a \gls{BS}, an \gls{RIS}, $K$ single-antenna users in the set $\mathcal{K}$, and $L$ point-targets in the set $\mathcal{L}$. The \gls{BS} is equipped with an $N_t$-element transmit array and an $N_r$-element receive array, with antenna positions $\myvec{p}_t(n_t), \myvec{p}_r(n_r) \in \mathbb{R}^3$ and corresponding normal vectors $\myvec{w}_t, \myvec{w}_r$ defining the main radiation directions. The \gls{RIS} contains $N_s$ passive elements, with the $n_s$-th element's position and normal vector being $\myvec{p}_s(n_s)$ and $\myvec{w}_s$, respectively. The position of any user or target $a \in \mathcal{K} \cup \mathcal{L}$ is $\myvec{p}_a$. To model potential blockages, an indicator $\alpha_a \in \{0, 1\}$ describes the \gls{LoS} status ($\alpha_a=1$ for an available path, $0$ otherwise). All entities operate within the \gls{BS}'s near-field region, bounded by the Rayleigh distance $R_{\text{RL}} = 2D^2/\lambda$, where $D$ is the \gls{BS} antenna aperture and $\lambda$ is the signal wavelength.
	
	
	\vspace{-2ex}
	\subsection{Transmitted ISAC Signal}
	\vspace{-2ex}
	Under a fully-digital beamforming architecture \cite{hua2023optimal}, the signal transmitted from the \gls{BS} at time $t$ ($t = 1,\dots, T$, where $T$ is the block length) is given by $\myvec{x}[t] = \sum_{k \in \mathcal{K}} \myvec{f}_k c_k[t] + \myvec{s}[t]$,
	where $\myvec{f}_k \in \mathbb{C}^{N_t}$ is the communication beamforming vector for user $k$, and $c_k[t]$ is the corresponding data symbol, satisfying $\myexpect{c_k[t]c_l^*[t]} = \delta[k-l]$. The term $\myvec{s}[t] \in \mathbb{C}^{N_t}$ represents the dedicated sensing signal with covariance $\mymat{R}_s = \myexpect{\myvec{s}[t]\myvec{s}^H[t]}$. The total transmit covariance is:
	\begin{equation} \label{eqn:signal_cov_detailed}
		\mymat{R}_x = \myexpect{\myvec{x}[t]\myvec{x}^H[t]} = \sum\nolimits_{k\in\mathcal{K}} \myvec{f}_k \myvec{f}_k^H + \mymat{R}_s.
	\end{equation}
	\vspace{-2ex}
	\subsection{Communication Model}
	\vspace{-2ex}
	For user $k \in \mathcal{K}$, the overall downlink channel $\myvec{h}_{t,k}(\mymat{\Phi}) \in \mathbb{C}^{N_t}$ is a sum of the \gls{LoS} path (if it exists) and the \gls{BS}-\gls{RIS}-user path:
	\begin{equation} \label{eqn:comm_channel_final_detailed}
		\myvec{h}_{t,k}(\mymat{\Phi}) = {\alpha_k \myvecb{h}_{t,k}}
		+ {\left(\myvec{h}_{s,k}^T \mymat{\Phi} \mymat{G}_t\right)^T}.
	\end{equation}
	Following the near-field channel model \cite{zhang2022beam}, the components of the \gls{LoS} channel, $\myvecb{h}_{t,k}\in \mathbb{C}^{N_t}$, are defined as:
	\begin{equation}
		\left[\myvecb{h}_{t,k}\right]_{n_t} = \sqrt{F(\myvec{p}_k - \myvec{p}_t(n_t), \myvec{w}_t)} \beta(\myvec{p}_k - \myvec{p}_t(n_t)),
	\end{equation}
	where $F(\myvec{p}_k-\myvec{p}_t(n_t),\myvec{w}_t)$ is the radiation pattern, given by: 
	\begin{equation*}
		F(\myvec{p},\!\myvec{w}) \!
		= \!
		\begin{cases}
			2(b\!+\!1) \cos^b(\psi(\myvec{p},\!\myvec{w}) ), & \psi(\myvec{p},\!\myvec{w}) \!\in\! [0, \pi/2], \\
			0, & \text{otherwise},
		\end{cases}
	\end{equation*}
	where $\psi(\myvec{p},\myvec{w})$ is the angle between the direction vector $\myvec{p}$ and the array's normal vector $\myvec{w}$. The parameter \( b \) determines the Boresight gain, which is 2 for the dipole. The term $\beta(\myvec{p}_k-\myvec{p}_t(i,l))$ denotes the attenuation and phase change caused by the path, with  $\beta(\myvec{p})$ being
	\begin{equation}
		\beta(\myvec{p}) = {\lambda}/\left({4\pi\norm{\myvec{p}}}\right)
		\cdot e^{-j\frac{2\pi}{\lambda} \norm{\myvec{p}}}.
	\end{equation}
	The \gls{BS}-\gls{RIS}-user path consists of three components: the BS-RIS channel $\mymat{G}_t \in \mathbb{C}^{N_s \times N_t}$, the RIS reflection matrix $\mymat{\Phi} \in \mathbb{C}^{N_s \times N_s}$, and the RIS-user channel $\myvec{h}_{s,k} \in \mathbb{C}^{N_s}$. Their elements are:
	\begin{align}
		\begin{split}
			[\mymat{G}_t]_{n_s, n_t} \! = & \beta(\myvec{p}_s\!(n_s) \!-\! \myvec{p}_t(n_t)) \!\sqrt{\!F\!(\myvec{p}_s(n_s) \!-\! \myvec{p}_t(n_t),\! \myvec{w}_t)} \\
			& \times \sqrt{F(\myvec{p}_t(n_t) \!-\! \myvec{p}_s(n_s), \myvec{w}_s)},
		\end{split} \label{eqn:BS_RIS_channel} \\
		\mymat{\Phi} = & \diag(e^{j\theta_1}, \dots, e^{j\theta_{N_s}}), \label{eqn:ris_response} \\
		\left[\myvec{h}_{s,k}\right]_{n_s} = & \sqrt{F(\myvec{p}_k - \myvec{p}_s(n_s), \myvec{w}_s)} \beta(\myvec{p}_k - \myvec{p}_s(n_s)), \label{eqn:RIS_user_channel}
	\end{align}
	where $\theta_{n_s} (n_s = 1, \dots, N_s)$ is the response of the $n_s$-th \gls{RIS} element. The received signal at user $k$ is
	\begin{equation}
		y_k[t] = \myvec{h}_{t,k}^T(\mymat{\Phi})\myvec{x}[t] + n_k[t],
	\end{equation}
	where $n_k[t] \sim \mathcal{CN}(0,\sigma_k^2)$ is the additive noise. Then, the \gls{SINR} is:
	\begin{equation} \label{eqn:comm_SINR}
		\eta_k = \frac{\abs{\myvec{h}_{t,k}^T(\mymat{\Phi}) \myvec{f}_k}^2}{\sum_{j \neq k} \abs{\myvec{h}_{t,k}^T(\mymat{\Phi}) \myvec{f}_j}^2 + \myvec{h}_{t,k}^T(\mymat{\Phi}) \mymat{R}_s \myvec{h}_{t,k}^*(\mymat{\Phi}) + \sigma_k^2}.
	\end{equation}
	The achievable communication rate for user \( k \) 
	 is used as the performance metric, given by 
	 \cite{zhang2022beam}:
	\begin{equation} \label{eqn:shannon_capacity}
		R_k = \log_2 (1 + \eta_k).
	\end{equation}
	\vspace{-3ex}
	\subsection{Sensing Model}
	\vspace{-2ex}
	The sensing channel constitutes a round-trip path, consisting of the forward path from the BS to a target and the backward path from the target to the BS. Similarly to \eqref{eqn:comm_channel_final_detailed}, the forward and backward channels, denoted as $\myvec{h}_{t,l}(\mymat{\Phi}) \in \mathbb{C}^{N_t}$ and $\myvec{h}_{l,r}(\mymat{\Phi}) \in \mathbb{C}^{N_r}$, are composed of the direct path (if it exists) and the \gls{RIS}-assisted path:
	\begin{equation*}
		\myvec{h}_{t,l}(\mymat{\Phi}) \!=\! \alpha_l \myvecb{h}_{t,l} + \left(\myvec{h}_{s,l}^T \mymat{\Phi} \mymat{G}_t\right)^T\!\!, 
		\myvec{h}_{l,r}(\mymat{\Phi}) \!=\! \alpha_l \myvecb{h}_{l,r} + \mymat{G}_r \mymat{\Phi} \myvec{h}_{s,l}.
	\end{equation*}
	The components of the forward and backward paths are the direct BS-to-target \gls{LoS} channel $\myvecb{h}_{t,l} \in \mathbb{C}^{N_t}$, the RIS-target channel $\myvec{h}_{s,l} \in \mathbb{C}^{N_s}$, the BS-RIS channel gain $\mymat{G}_t$ given in \eqref{eqn:BS_RIS_channel}, the target-to-BS \gls{LoS} channel  $\myvecb{h}_{l,r}\mathbb{C}^{N_r}$ and the RIS-BS-receive channel gain $\mymat{G}_r \in \mathbb{C}^{N_r \times N_s}$. The elements of these vectors or matrices are as follows: 
	\begin{align*}
		\left[\myvecb{h}_{t,l}\right]_{n_t} &= \sqrt{F(\myvec{p}_l - \myvec{p}_t(n_t), \myvec{w}_t)} \beta(\myvec{p}_l - \myvec{p}_t(n_t)),\\
		\left[\myvec{h}_{s,l}\right]_{n_s} & = \sqrt{F(\myvec{p}_l - \myvec{p}_s(n_s), \myvec{w}_s)} \beta(\myvec{p}_l - \myvec{p}_s(n_s)),\\
		[\myvecb{h}_{l,r}]_{n_r} &= \sqrt{F(\myvec{p}_r(n_r) - \myvec{p}_l, \myvec{w}_r)} \beta(\myvec{p}_r(n_r) - \myvec{p}_l), \\
		[\mymat{G}_r]_{n_r, n_s} &=  \beta(\myvec{p}_r(n_r) \!-\! \myvec{p}_s(n_s))\sqrt{F(\myvec{p}_r(n_r) \!-\! \myvec{p}_s(n_s),\! \myvec{w}_s)} \\
		& \quad \times \sqrt{F(\myvec{p}_s(n_s)\! -\! \myvec{p}_r(n_r),\! \myvec{w}_r)}.
	\end{align*}
	The complete round-trip sensing channel matrix for target $l$ is 
	\begin{equation} \label{eqn:sense_channel_final}
		\mymat{H}_l(\mymat{\Phi}) = \gamma_l \myvec{h}_{l,r}(\mymat{\Phi}) \myvec{h}_{t,l}^T(\mymat{\Phi}),
	\end{equation}
	where $\gamma_l$ is the target's \gls{RCS}. The BS's total received echo signal is
	\begin{equation}
		\myvec{y}_r[t] = \sum\nolimits_{l \in \mathcal{L}} \mymat{H}_l(\mymat{\Phi}) \myvec{x}[t] + \myvec{n}_r[t],
	\end{equation}
	where $\myvec{n}_r[t] \sim \mathcal{CN}(\mathbf{0}, \sigma_r^2 \mymat{I}_{N_r})$ is the sensing receiver's noise. For multi‐target MIMO sensing, the transmit covariance $\mymat{R}_x$ must be shaped to achieve two complementary objectives \cite{liu2020joint,meng2024multi,stoica2007probing}:
	1)~\textbf{Beam‐pattern gain}: maximize the power delivered to each target $l$, quantified by
	\(
	\myvec{h}_{t,l}^T \,\mymat{R}_x\, \myvec{h}_{t,l}^*;
	\)
	2)~\textbf{Cross‐correlation suppression}: minimize mutual interference between any two targets $l\neq l'$, quantified by $\bigl|\myvec{h}_{t,l}^T \,\mymat{R}_x\, \myvec{h}_{t,l'}^*\bigr|$.	

	Our key innovation lies in extending these classic sensing metrics to the near-field. While far-field designs optimize beampattern and cross-correlation over angles only \cite{stoica2007probing,liu2020joint,meng2024multi}, our near-field formulation enforces these criteria over a joint angle-and-range space. This enables resolving co-directional targets. Furthermore, unlike other common metrics such as the \gls{SINR} or \gls{MI} \cite{galappaththige2024near,hua2025near}, this approach is agnostic to the targets' often unknown \glspl{RCS}, making it more robust for practical deployments. More critically, optimizing for SINR or beampattern gain without suppressing cross-correlation leads to highly correlated target echoes, making classical indication techniques like Capon fail to resolve multiple targets \cite{stoica2007probing, meng2024multi}.
	
	\section{Joint Beamforming and RIS Design}
	\label{sec:bf_opt}
	\subsection{Optimization formulation}
	\vspace{-2ex}
	We aim to jointly design the BS transmit beamformers $\{\myvec{f}_k\}$, the sensing covariance $\mymat{R}_s$, and the RIS phase shifts $\mymat{\Phi}$ to optimize the performance of \gls{ISAC} system, which is formulated as follows:
	\begin{subequations} \label{eqn:bf_opt_RIS}
		\begin{align}
			\max_{\substack{\{\myvec{f}_k\}, \mathbf{R}_s\succeq 0, \\ \mymat{R}_x, \mymat{\Phi}, \mu, \{\theta_{n_s}\}}} \quad & \mu \label{eqn:opt_obj} \\
			\text{s.t.} \quad & w_l \left(\myvec{h}_{t,l}^T(\mymat{\Phi}) \mymat{R}_x \myvec{h}_{t,l}^*(\mymat{\Phi})\right) \geq \mu, \forall l \in \mathcal{L}, \label{eqn:opt_const_sensing_bp_ris}\\
			& w_{l,l'} \!\abs{\myvec{h}_{t,l}^T(\!\mymat{\Phi}\!) \mymat{R}_x \myvec{h}_{t,l'}^*(\!\mymat{\Phi}\!)} \!\leq\! \varepsilon \mu,  \forall l \!\neq\! l', \label{eqn:opt_const_cross_ris}\\
			& R_k(\{\myvec{f}_k\}, \mymat{R}_s, \mymat{\Phi}) \geq R_{\min,k},  \forall k \in \mathcal{K}, \label{eqn:opt_const_comm_ris}\\
			& \operatorname{Tr}(\mathbf{R}_x) \leq P_{\max}, \label{eqn:opt_const_bs_power_ris} \\
			& \mymat{R}_x = \sum\nolimits_{k\in\mathcal{K}} \myvec{f}_k \myvec{f}_k^H +\mymat{R}_s, \label{eqn:opt_const_cov_ris} \\
			& \mymat{\Phi} = \text{diag}(e^{j\theta_1}, \dots, e^{j\theta_{N_s}}).
			\label{eqn:opt_const_ris_phase}
		\end{align}
	\end{subequations}
	The optimization in \eqref{eqn:bf_opt_RIS} maximizes an auxiliary variable $\mu$, which represents the worst-case sensing beampattern gain weighted by $w_l$ across all targets (objective \eqref{eqn:opt_obj}), subject to a series of constraints. Specifically, each target's weighted beampattern gain must be at least $\mu$ \eqref{eqn:opt_const_sensing_bp_ris}; the magnitude of the cross-correlation between any two targets must not exceed a fraction $\varepsilon$ of $\mu$ to suppress inter-target interference \eqref{eqn:opt_const_cross_ris}; each user's communication rate must meet a minimum requirement $R_{\min,k}$ \eqref{eqn:opt_const_comm_ris}; the total transmit power is limited by $P_{\max}$ \eqref{eqn:opt_const_bs_power_ris}; the equations of the overall transmit covariance $\mymat{R}_x$ \eqref{eqn:opt_const_cov_ris} and \gls{RIS} phase shifts \eqref{eqn:opt_const_ris_phase} are repetitions of \eqref{eqn:signal_cov_detailed} and \eqref{eqn:ris_response}. 
	
	The proposed framework in \eqref{eqn:bf_opt_RIS} is uniquely tailored for near-field multi-target ISAC, and its novelty is twofold.  First, the explicit inclusion of the cross-correlation suppression constraint \eqref{eqn:opt_const_cross_ris} is critical. It ensures that the waveforms reflected from different targets are nearly orthogonal, which is critical for multi-target indication \cite{stoica2007probing}. Second, the power of this framework is unlocked by using near-field channels. Because the near-field channel vectors \(\myvec{h}_{t,l}(\mymat{\Phi})\) capture range-dependent information, allowing the beampattern and cross-correlation constraints to be enforced over a joint angle-and-range space, which is the core mechanism that enables our design to resolve co-directional users and targets.

	This problem is non-convex and challenging to solve due to the coupling between the optimization variables $\mymat{R}_x$ and $\mymat{\Phi}$, the unit-modulus constraint on the RIS phases \eqref{eqn:opt_const_ris_phase}, and the SINR constraints \eqref{eqn:opt_const_comm_ris} with respect to the beamforming vectors $\{\myvec{f}_k\}$. To tackle this, we adopt an \gls{AO} framework. We decouple the problem by first optimizing the BS transmit covariance $\mymat{R}_x$ while keeping the RIS phase shifts $\mymat{\Phi}$ fixed. Subsequently, we optimize $\mymat{\Phi}$ using the updated $\mymat{R}_x$. This two-step process is repeated until convergence.
	\vspace{-2ex}
	\subsection{Fix $\mymat{\Phi}$, Optimize BS Beamforming}
	\vspace{-2ex}
	When $\mymat{\Phi}$ is fixed, the channels $\myvec{h}_{t,a}(\mymat{\Phi})$ become constant vectors. The problem \eqref{eqn:bf_opt_RIS} reduces to an optimization over the BS beamformers $\{\myvec{f}_k\}$ and the sensing covariance $\mymat{R}_s$. This subproblem is still non-convex due to the communication rate constraint \eqref{eqn:opt_const_comm_ris} and the quadratic covariance definition \eqref{eqn:opt_const_cov_ris}. This is tackled using \gls{SDR}.
	
	First, using the Shannon capacity formula \eqref{eqn:shannon_capacity} and the definition of \gls{SINR} \eqref{eqn:comm_SINR}, \eqref{eqn:opt_const_comm_ris} is equivalent to \cite{wang2023near}:
	\begin{equation}
		\xi_{k} \abs{\myvec{h}_{t,k}^T(\mymat{\Phi}) \myvec{f}_k}^2 \geq \myvec{h}_{t,k}^T(\mymat{\Phi}) \mymat{R}_x \myvec{h}_{t,k}^*(\mymat{\Phi}) + \sigma_k^2,
	\end{equation}
	where $\xi_{k} =
	{2^{\,R_{\min,k}}}/({2^{R_{\min,k}} - 1})$. We then introduce variables $\mymat{F}_k = \myvec{f}_k \myvec{f}_k^H$, and relax the rank-one constraints $\text{rank}(\mymat{F}_k)=1$. The subproblem becomes the following \gls{SDP}: 
	\begin{subequations} \label{eqn:sdr_subproblem}
		\begin{align}
			\max_{\substack{\{\mymat{F}_k\succeq 0\},\mathbf{R}_s\succeq 0, \mu}} \quad & \mu \\
			\text{s.t.} \quad & \text{Constraints } \eqref{eqn:opt_const_sensing_bp_ris}, \eqref{eqn:opt_const_cross_ris}, \eqref{eqn:opt_const_bs_power_ris} \\
			& \text{Tr}\left(\mymat{H}_{k}(\mymat{\Phi})(\xi_k\mymat{F}_k - \mymat{R}_x)\right) \geq \sigma_k^2, \forall k \\
			& \mymat{R}_x = \sum\nolimits_{k\in\mathcal{K}} \mymat{F}_k +\mymat{R}_s,
		\end{align}
	\end{subequations}
	where $\mymat{H}_{k}(\mymat{\Phi}) = \myvec{h}_{t,k}^*(\mymat{\Phi})\myvec{h}_{t,k}^T(\mymat{\Phi})$. This \gls{SDP} can be solved efficiently using standard convex solvers. From the optimal solution $\{\mymat{F}_k^*\}_{k \in \mathcal{K}}$, we can construct an optimal beaming vector $\myvech{f}_k$ without loss of optimality as follows \cite{liu2020joint}:
	\begin{equation} \label{eqn:rank_one_recovery}
		\myvech{f}_k = \left(\myvec{h}_{t,k}^T(\mymat{\Phi}) \mymat{F}_k^* \myvec{h}_{t,k}^*(\mymat{\Phi})\right)^{-1/2} \mymat{F}_k^* \myvec{h}_{t,k}^*(\mymat{\Phi}).
	\end{equation}
	The optimality of this construction is justified below.
	
	\vspace{-3ex}
	\begin{proof}
		Let the solution to the \gls{SDP} in \eqref{eqn:sdr_subproblem} be $\{\mymat{F}_k^*\}$ and $\mymat{R}_s^*$. We construct the beamformers $\{\myvech{f}_k\}$ using \eqref{eqn:rank_one_recovery} and define a new sensing covariance matrix $\mymath{R}_s = \mymat{R}_s^* + \sum_{k\in\mathcal{K}}(\mymat{F}_k^* - \myvech{f}_k\myvech{f}_k^H)$. According to \cite[Theorem 1]{liu2020joint}, the constructed vectors satisfy $\myvech{f}_k\myvech{f}_k^H \preceq \mymat{F}_k^*$, which ensures that $\mymath{R}_s \succeq \mathbf{0}$. Since the total transmit covariance matrix remains unchanged, the sensing constraints \eqref{eqn:opt_const_sensing_bp_ris}, the cross-correlation constraints \eqref{eqn:opt_const_cross_ris}, and the power constraint \eqref{eqn:opt_const_bs_power_ris} are still satisfied. Furthermore, the construction in \eqref{eqn:rank_one_recovery} guarantees that $\abs{\myvec{h}_{t,k}^T(\mymat{\Phi})\myvech{f}_k}^2 = \myvec{h}_{t,k}^T(\mymat{\Phi})\mymat{F}_k^*\myvec{h}_{t,k}^*(\mymat{\Phi})$ \cite[Theorem 1]{liu2020joint}. This ensures that the left-hand side of the SINR constraint in \eqref{eqn:sdr_subproblem} remains unchanged, meaning the communication rate constraint \eqref{eqn:opt_const_comm_ris} also holds. Therefore, the constructed rank-one solution $\{\myvech{f}_k\}$ and $\mymath{R}_s$ is feasible and achieves the same optimal objective value $\mu$.
	\vspace{-3ex}
	\end{proof}
	\vspace{-2ex}
	\subsection{Fix BS Beamforming, Optimize $\mymat{\Phi}$}
	\vspace{-2ex}
	When the BS transmit covariance $\mymat{R}_x$ is fixed, the original problem \eqref{eqn:bf_opt_RIS} reduces to a subproblem over the RIS phase shifts $\mymat{\Phi}$ and the objective value $\mu$.
	This subproblem is non-convex due to the unit-modulus constraint \eqref{eqn:opt_const_ris_phase}, which will be solved with \gls{SDR}. First, let us define a vector $\myvec{\phi} = [e^{j\theta_1}, \dots, e^{j\theta_{N_s}}]^T$ containing the RIS phase shifts, such that $\mymat{\Phi} = \diag(\myvec{\phi})$. The channel from the BS to any user or target $a \in \mathcal{K} \cup \mathcal{L}$ can be expressed as an affine function of $\myvec{\phi}$:
	\begin{equation}
		\myvec{h}_{t,a}(\myvec{\phi}) = \alpha_a \myvecb{h}_{t,a} + \mymat{A}_a \myvec{\phi},
	\end{equation}
	where $\mymat{A}_a = \mymat{G}_t^T \diag(\myvec{h}_{s,a})$. 
	Then, we define an augmented vector $\myvecb{\phi} = [\myvec{\phi}^T, 1]^T$ and a new matrix variable $\mymatb{\Psi} = \myvecb{\phi}\myvecb{\phi}^H \in \mathbb{C}^{(N_s+1)\times(N_s+1)}$. 
	Thus, any quadratic form of $\myvec{\phi}$ becomes a linear function of $\mymatb{\Psi}$. The beampattern and cross-correlation terms can be expressed as $\text{Tr}(\mymat{M}_{l,l'} \mymatb{\Psi})$, with $\mymat{M}_{l,l'}$ defined as:
	\begin{equation*}
		\mymat{M}_{l,l'} =
		\begin{pmatrix}
			\mymat{A}_l^H \mymat{R}_x^T \mymat{A}_{l'} & \mymat{A}_l^H \mymat{R}_x^T (\alpha_{l'} \myvecb{h}_{t,l'}) \\
			(\alpha_l \myvecb{h}_{t,l})^T \mymat{R}_x \mymat{A}_{l'}^* & (\alpha_l \myvecb{h}_{t,l})^T \mymat{R}_x (\alpha_{l'} \myvecb{h}_{t,l'})^*
		\end{pmatrix}.
	\end{equation*}
	
	Similarly, the SINR constraint for user $k$ can be rewritten as the linear inequality $\text{Tr}(\mymat{M}_{k}^{\text{SINR}} \mymatb{\Psi}) \geq \Gamma_k \sigma_k^2$, where $\mymat{M}_{k}^{\text{SINR}}$ is constructed by first defining $\mymat{S}_k = \myvec{f}_k \myvec{f}_k^H - \Gamma_k (\sum_{j \neq k} \myvec{f}_j \myvec{f}_j^H + \mymat{R}_s)$, and then:
	\begin{equation*}
		\mymat{M}_{k}^{\text{SINR}} =
		\begin{pmatrix}
			\mymat{A}_k^H \mymat{S}_k^T \mymat{A}_{k} & \mymat{A}_k^H \mymat{S}_k^T (\alpha_{k} \myvecb{h}_{t,k}) \\
			(\alpha_k \myvecb{h}_{t,k})^T \mymat{S}_k^* \mymat{A}_{k}^* & (\alpha_k \myvecb{h}_{t,k})^T \mymat{S}_k^* (\alpha_{k} \myvecb{h}_{t,k})^*
		\end{pmatrix}.
	\end{equation*}
	
	After we drop the rank-one constraint on $\mymatb{\Psi}$, and relax the unit-modulus constraint $|\phi_{n_s}|^2=1$ into 
	$[\mymatb{\Psi}]_{n_s,n_s} = 1$ for all $n_s=1, \dots, N_s+1$, 
	the relaxed subproblem for optimizing the RIS phase shifts becomes the following convex \gls{SDP}:
	\begin{subequations} \label{eqn:ris_sdr_subproblem}
		\begin{align}
			\max_{\mymatb{\Psi} \succeq 0, \mu} \quad & \mu \\
			\text{s.t.} \quad & w_l \text{Tr}(\mymat{M}_{l,l} \mymatb{\Psi}) \geq \mu, \quad \forall l \in \mathcal{L}, \\
			& w_{l,l'} \abs{\text{Tr}(\mymat{M}_{l,l'} \mymatb{\Psi})} \leq \varepsilon \mu, \quad \forall l \neq l', \\
			& \text{Tr}(\mymat{M}_{k}^{\text{SINR}} \mymatb{\Psi}) \geq \Gamma_k \sigma_k^2, \quad \forall k \in \mathcal{K}, \\
			& [\mymatb{\Psi}]_{n_s, n_s} = 1, \quad n_s = 1, \dots, N_s+1.
		\end{align}
	\end{subequations}
This \gls{SDP} can be solved efficiently using convex optimization solvers. If the resulting optimal matrix $\mymatb{\Psi}^*$ is not rank-one, a rank-one solution for $\myvec{\phi}$ is recovered using a deterministic method. First, we select the dominant eigenvector of $\mymatb{\Psi}^*$, denoted as $\myvecb{v}$, and normalize it by its last element to obtain a candidate vector, i.e., $\myvecb{\phi}_{\text{cand}} = {\myvecb{v}}/({[\myvecb{v}]_{N_s+1}})$. Then, the final RIS phase-shift vector $\myvec{\phi}$ is formed by taking the first $N_s$ elements of $\myvecb{\phi}_{\text{cand}}$ and ensuring they have unit modulus, i.e., $	\phi_{n_s} = {[\myvecb{\phi}_{\text{cand}}]_{n_s}}/{\abs{[\myvecb{\phi}_{\text{cand}}]_{n_s}}},\text{for } n_s=1, \dots, N_s.$

\section{Simulation Results}
\label{sec:result}

In this section, we evaluate the performance of our proposed joint beamforming and RIS optimization framework. We present numerical results for a near-field scenario with communication users and sensing targets in \gls{NLoS} positions, demonstrating the effectiveness of the RIS in establishing virtual links for multiple users and targets.
\vspace{-5ex}
\subsection{Parameter Configuration}
\vspace{-2ex}

We consider an \gls{ISAC} system at a carrier frequency of $f_c = 30$ GHz, with a wavelength of $\lambda = 10$ mm. The maximum transmit power at the \gls{BS} is $P_{\max} = 23$ dBm. The noise power is set to -80 dBm for communication users and -110 dBm for the sensing receiver. The \gls{BS} is equipped with $7 \times 7$ \gls{UPA}s for both transmission ($N_t=49$) and reception ($N_r=49$). The \gls{RIS} is also a $7 \times 7$ \gls{UPA} ($N_s=49$). The inter-element spacing for all arrays is $\lambda/2$, resulting in a Fraunhofer distance of approximately $0.49$ m.

The geometry is configured to create a challenging blockage scenario. The \gls{BS} is at $[0, -0.3, 0.3]$ m and the \gls{RIS} is at the origin. An obstacle blocks the \gls{LoS} path for any location with a y-coordinate $y \ge 0$. The system serves $K=3$ users and $L=3$ targets. User 1 is an \gls{LoS} user at $[0, -0.07, 0.07]$ m. Users 2 and 3 are \gls{NLoS} users at $[0, 0.049, 0.074]$ m and $[0, 0.15, 0.22]$ m. All three targets are in \gls{NLoS} positions at $[0, 0.049, 0.049]$ m, $[0, 0.12, 0.12]$ m, and $[0, 0.098, 0.196]$ m. Notably, some \gls{NLoS} users (User 2 and 3) and targets (Target 1 and 2) are co-directional relative to the \gls{RIS}, creating a demand for near-field separation. The minimum rate requirements are $14$, $7$, and $7$ bps/Hz for the three users. The cross-correlation tolerance is set to $\varepsilon = 0.1$. The block length is $T = 1000$.

To evaluate performance, our proposed method is compared against two benchmarks. The first is a conventional \textbf{\gls{FFBF}} design, which replaces the near-field spherical wave model with a planar wave model.  Additionally, since far‐field users or targets at the same direction become fully coherent, we relax the communication constraint by lowering $R_{\min}$ to 0.95 bps/Hz and remove  the cross-correlation constraint \eqref{eqn:opt_const_cross_ris} for the target pair 
$(1,2)$, thereby ensuring problem feasibility. The second, termed \textbf{\gls{NCCS}}, utilizes our proposed algorithm but without the inter-target cross-correlation constraint \eqref{eqn:opt_const_cross_ris}. This comparison serves to highlight the respective benefits of near-field focusing and explicit cross-correlation suppression.

\begin{figure}[htb]
	\centering
	\subfloat[SINR map with proposed method]{
		\includegraphics[width=0.9\linewidth]{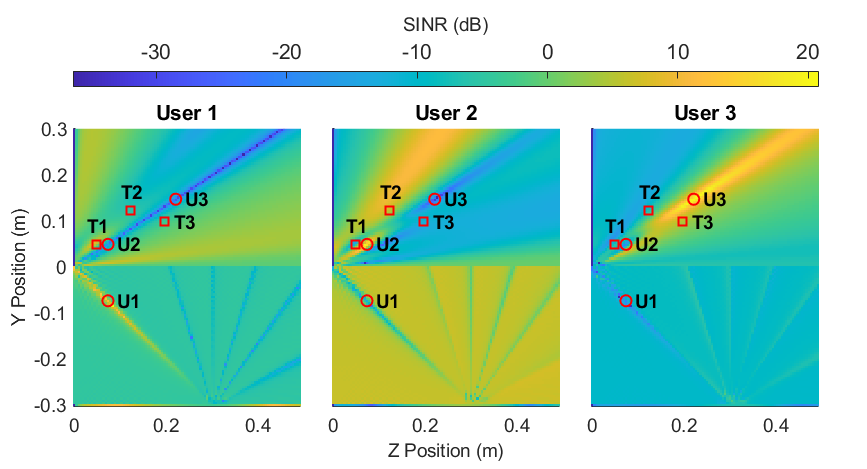} 
		\label{fig:sinr_proposed}
	}
	\vspace{-0.3cm} 
	\subfloat[SINR map with FFBF]{
		\includegraphics[width=0.9\linewidth]{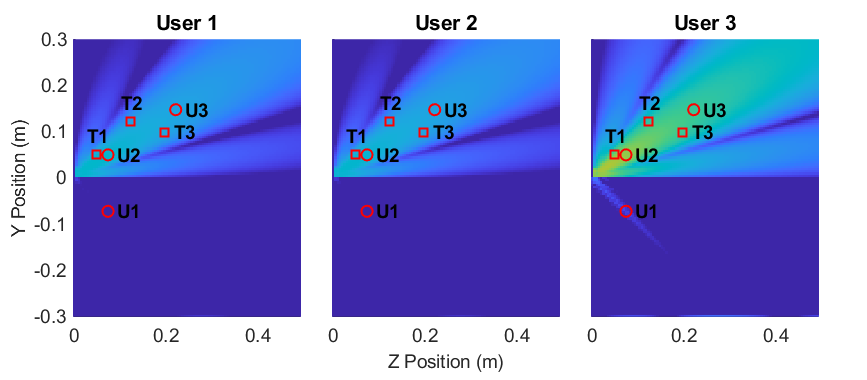}
		\label{fig:sinr_ffbf}
	}
	\caption{\Gls{SINR} distribution for communications users.}
	\vspace{-0.5cm}
	\label{fig:comm_sinr_ris}
\end{figure}
\vspace{-2ex}
\subsection{Communication Performance}
\vspace{-2ex}
We first evaluate the communication performance. Fig.~\ref{fig:comm_sinr_ris} illustrates each user's \gls{SINR} distribution across the service area. As shown in Fig.~\ref{fig:sinr_proposed}, our proposed near-field design forms a strong, focused beam directly towards the \gls{LoS} user (User 1), resulting in a high \gls{SINR} at its location. More importantly, for the \gls{NLoS} users (Users 2 and 3), our joint optimization routes the signal via the \gls{RIS}, successfully bypassing the obstacle and creating robust virtual \gls{LoS} links. This results in sharp \gls{SINR} peaks at the \gls{NLoS} users' locations, enabling reliable high-rate communication where it would otherwise be impossible. In contrast, Fig.~\ref{fig:sinr_ffbf} shows the performance of a conventional \gls{FFBF} benchmark. The inability of far-field model to perform distance-aware focusing prevents it from effectively leveraging the \gls{RIS} to serve the \gls{NLoS} users at the same direction of \gls{RIS}, leading to significantly degraded performance in the blocked region.

\begin{figure}[htb]
	\centering
	\includegraphics[width=0.9\linewidth]{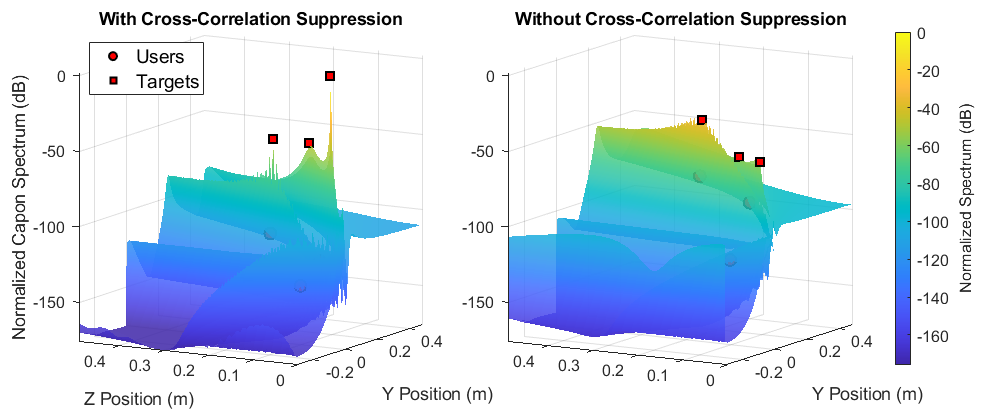} 
	\caption{Normalized Capon spectrum for sensing, comparing the proposed method with cross-correlation suppression (Left) against NCCS without suppression (Right).}
	\vspace{-0.6cm}
	\label{fig:sensing_capon_ris}
\end{figure}

\vspace{-2ex}
\subsection{Sensing Performance}
\vspace{-2ex}
Next, we assess the multi-target sensing capability. 
Fig.~\ref{fig:sensing_capon_ris} compares the normalized Capon spectrum of our proposed method (left) against NCCS without cross-correlation suppression (right). Our design 
results in three distinct peaks that precisely match the true locations of the \gls{NLoS} targets, confirming that our near-field beamformer can separates targets along the same direction to the \gls{RIS} (Targets 1 and 2). In contrast, the \gls{NCCS} benchmark fails to resolve the targets effectively. Without the suppression constraint, the echoes from the different targets are highly correlated, leading to a smeared spectrum with significant sidelobes and ambiguous peaks. This result illustrates that the explicit suppression of cross-correlation is critical for enabling reliable multi-target indication in \gls{NLoS} scenarios.

	\vspace{-0.2cm}
	\section{Conclusion}
	In this paper, we addressed the critical challenge of \gls{LoS} blockage in near-field multi-target \gls{ISAC} systems. 
	We formulated a joint transmission beamforming and RIS phase shifts optimization problem to maximize sensing performance while guaranteeing communication quality and suppressing interference. An efficient \gls{AO} algorithm was developed to solve this complex non-convex problem. Simulation results confirm that our RIS-assisted approach can successfully bypass obstacles, providing robust and high-performance communication and sensing functionalities in challenging \gls{NLoS} environments. 

	\section{Acknowledgements}
	This research was supported by the European Research Council (ERC) under the European Union’s Horizon 2020 research and innovation program (grant No. 101000967), by the Israel Science Foundation (grant No. 536/22), and by the Manya Igel Centre for Biomedical Engineering and Signal Processing.
	
	\bibliographystyle{IEEEtran}
	\bibliography{2D}
	
\end{document}